\pdfoutput=1
\documentclass[a4paper]{article}
\usepackage{INTERSPEECH2020}
\usepackage{float}
\usepackage{amsmath,graphicx,amssymb}
\usepackage[dvipsnames]{xcolor}
\usepackage{tikz}
\usetikzlibrary{decorations.text}
\usetikzlibrary{calc}

\newcommand{\norm}[1]{\left\lVert#1\right\rVert}
\makeatletter
\newcommand\notsotiny{\@setfontsize\notsotiny\@vipt\@viipt}
\makeatother

\title{End-to-end Domain-Adversarial Voice Activity Detection}

\name{Marvin Lavechin $^{1*}$ \; Marie-Philippe Gill $^{2}$ \; Ruben Bousbib $^{1}$ \\ Herv\'{e} Bredin $^{3*}$ \; Leibny Paola Garcia-Perera $^{4}$}

\address{
$^{1}$ Cognitive Machine Learning team, \'{E}cole Normale Sup\'{e}rieure/INRIA, PSL, Paris, France\\
$^{2}$ \'{E}cole de Technologie Sup\'{e}rieure, Universit\'{e} du Qu\'{e}bec, Montreal, Canada \\
$^{3}$ LIMSI, CNRS, Univ. Paris-Sud, Universit\'{e} Paris-Saclay, Orsay, France \\
$^{4}$ Center for Language and Speech Processing, The Johns Hopkins University, Baltimore, USA}

\email{$^{*}$marvinlavechin@gmail.com, bredin@limsi.fr}

\begin{document}

\maketitle

\begin{abstract}
Voice activity detection is the task of detecting speech regions in a given audio stream or recording. First, we design a neural network combining trainable filters and recurrent layers to tackle voice activity detection directly from the waveform. Experiments on the challenging DIHARD dataset show that the proposed end-to-end model reaches state-of-the-art performance and outperforms a variant where trainable filters are replaced by standard cepstral coefficients. Our second contribution aims at making the proposed voice activity detection model robust to domain mismatch. To that end, a domain classification branch is added to the network and trained in an adversarial manner. The same DIHARD dataset, drawn from 11 different domains is used for evaluation under two scenarios. In the \emph{in-domain} scenario where the training and test sets cover the exact same domains, we show that the domain-adversarial approach does not degrade performance of the proposed end-to-end model. In the \emph{out-domain} scenario where the test domain is different from training domains, it brings a relative improvement of more than 10\%. Finally, our last contribution is the provision of a fully reproducible open-source pipeline than can be easily adapted to other datasets.
\end{abstract}

\noindent\textbf{Index Terms}: voice activity detection, domain adversarial training, sincnet, long short-term memory

\section{Introduction and related work}
\label{sec:intro}

Voice activity detection is one of the earliest building blocks of every speech processing pipeline, such as speaker recognition and speaker diarization. Learnt embeddings, able to discriminate speech segments from non-speech segments in audio recordings, might be sensitive to domain mismatch and lead to errors and bias in the subsequent processing steps.

A major assumption in the supervised learning scenario is that both the training and test data must be unbiased samples of the same underlying distribution. A mismatch between the training and the test set can lead to a high performance discrepancy. When learning from multiple sources or domains, models can specialize themselves to perform particularly well on some of these domains, and poorly on some others. A viable strategy might consist of muting this domain information to improve robustness of learnt embeddings. Indeed, robustness against different factors such as domain, noise, reverberation or speaker is a key point for practical use and fruitful deployment of most of the automated speech analysis tools.

Approaches aiming at improving invariance and robustness of learnt features have been long studied. In \cite{seltzer2013investigation}, Seltzer and Yu show that noise-aware training improves performance on a speech recognition task. In \cite{ganin2015unsupervised}, Ganin and Lempitsky propose a domain adaptation approach to learn discriminative features from a labeled source domain. They ensure that these features are domain shift invariant, so that they can be applied to any unlabeled different (but related) domains. Such as ours, their architecture relies on a first branch responsible for solving the main task, and a secondary branch responsible for classifying the domain. The latter goes through a gradient reversal layer, forcing the network to extract domain-independent features.

In recent years, domain-adversarial approaches to solve such problems have gained interest and have been applied to a wide range of tasks \cite{Shinohara2016AdversarialML,hu2019adversarial,tu2019adversarial, sun_da_accent, Tripathi2018AdversarialLO,liu_end_to_end_adv_reco}. For instance, in \cite{Shinohara2016AdversarialML}, a primary task of senone classification and a secondary task of noise condition are jointly solved on an artificially noise-corrupted version of the Wall Street Journal dataset. A similar approach has been used in \cite{hu2019adversarial} where they study this domain-adversarial approach on the speech recognition task in a multilingual setup to extract language invariant representation. In \cite{tu2019adversarial}, they used the same approach for extracting speaker-invariant representation for the speech emotion recognition task.
However, to the best of our knowledge, it has never been shown that domain-adversarial training of DNNs could improve performances on the voice activity detection task.

In parallel, learning acoustic models directly from the raw waveform has been an active area of research \cite{Sainath2015LearningTS,zeghidour:hal-01888737,Ravanelli2018}. Our approach relies on the SincNet model \cite{Ravanelli2018} acting as a feature extractor. We show that such approaches are also sensitive to bias such as the domain, and muting this information can help the model to extract more robust features, and therefore improve performance on new unseen domains.

\begin{figure*}
\tikzstyle{element}=[rectangle,draw, font=\small]
\tikzstyle{arrow}=[->,thick]
\tikzstyle{label}=[right, font=\footnotesize]

\centering
\resizebox{\textwidth}{!}{%
    \begin{tikzpicture}
        \node[circle,draw,color=RoyalBlue,font=\small, inner sep=1pt, outer sep=0pt] (ly) at (8.5, 0.9) {$\mathcal{L}_v$};
        \node[circle] (ly2) at (1, 0.9) {};
        \node[circle,draw,color=BrickRed, font=\small, inner sep=1pt, outer sep=0pt] (ld) at (8.5, -2.05) {$\mathcal{L}_d$};
        \node[circle] (ld2) at (1, -2.05) {};
        \node[circle] (gradient_reversal) at (0.5,-0.6) {};
        \node[inner sep=0pt,text width=10mm] (audiotop) at (-1.9,0.9) {};
        \node[inner sep=0pt,text width=10mm] (audiobot) at (-1.9,-0
    .6) {};
        \node[inner sep=0pt,text width=10mm] (audio) at (-2,0)
                {\includegraphics[width=10mm]{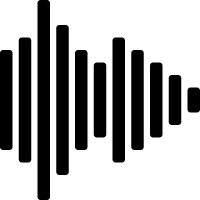}};
        \node[label] (speech) at (8.5, 0.15) {Speech};
        \node[label] (nospeech) at (8.5,-0.15) {Non-Speech};
        \node[label] (domain1) at (8.5, -0.55) {Domain 1};
        \node[label] (domain2) at (8.5, -0.8) {Domain 2};
        \node[label] (domaini) at (8.5, -1.05) {\hspace{0.4cm}\vdots};
        \node[label] (domainn) at (8.5, -1.45) {Domain n};
        \node[element] (sincnet) at (0,0) {SincNet};
        \node[element] (lstm1) at (2,0) {LSTM};
        \node[element] (lstm2) at (4,0) {LSTM};
        \node[element] (ff1) at (5.5,0) {FF};
        \node[element] (ff2) at (6.5,0) {FF};
        \node[element] (ff3) at (7.5,0) {FF};
        \node[element] (lstm3) at (2,-1) {LSTM};
        \node[element, align=center] (pool) at (5,-1) {Temporal Pooling};
        \node[element] (ff4) at (7.5, -1) {FF};
        \node[element, align=center, font=\notsotiny \linespread{0.4}] (grl) at (0.5,-1.5) {Gradient \\ Reversal};
        
        \draw[arrow] (audio) -- (sincnet);
        \draw[arrow] (sincnet) -- (lstm1) coordinate[midway] (aux){};
        \draw[arrow] (lstm1) -- (lstm2);
        \draw[arrow] (lstm2) -- (ff1);
        \draw[arrow] (ff1) -- (ff2);
        \draw[arrow] (ff2) -- (ff3);
        \draw[arrow] (lstm3) -- (pool);
        \draw[arrow] (pool) -- (ff4);
        \draw[arrow] (aux) |- (lstm3);
        \draw[thick] (ff3.east) -- ([xshift=0.4cm]ff3.east);
        \draw[thick] (ff4.east) -- ([xshift=0.4cm]ff4.east);
        \draw[arrow] ([xshift=0.4cm]ff3.east) |- (speech.west);
        \draw[arrow] ([xshift=0.4cm]ff3.east) |- (nospeech.west);
        \draw[arrow] ([xshift=0.4cm]ff4.east) |- (domain1.west);
        \draw[arrow] ([xshift=0.4cm]ff4.east) |- (domain2.west);
        \draw[arrow] ([xshift=0.4cm]ff4.east) |- ([yshift=-0.1cm]domaini.west);
        \draw[arrow] ([xshift=0.4cm]ff4.east) |- (domainn.west);

        \draw[arrow, color=RoyalBlue] (ly) -> node [text width=2.5cm,midway,above] {$\frac{\partial \mathcal{L}_v}{\partial \theta_v}$} (ly2);
        \draw[arrow, color=BrickRed] (ld) -| node [text width=-6cm,midway,below] {$\frac{\partial \mathcal{L}_d}{\partial  \theta_d }$} (grl);
        
        \def\myshift#1{\raisebox{1ex}};
        \draw[arrow, color=ForestGreen] (grl.north) |- node [midway,below,text width=3.5cm] {$-\lambda \frac{\partial \mathcal{L}_d }{\partial  \theta_f }$} (audiobot);
        \draw[arrow, color=ForestGreen] (ly2) -> node [text width=0.5cm,midway,above] {$\frac{\partial \mathcal{L}_v }{\partial  \theta_f}$} (audiotop);
        
        \def\myshift#1{\raisebox{-1.8ex}};
        
        \draw[RoyalBlue,thick, draw] ($(lstm1.south west)+(-0.3,0.9)$)  rectangle ($(ff3.north east)+(0.3,-0.6)$);
        \draw[BrickRed,thick, draw] ($(lstm3.south west)+(-0.3,0.6)$)  rectangle ($(ff4.north east)+(0.3,-1)$);
        \draw[ForestGreen,thick, draw] ($(sincnet.south west)+(-0.8,0.9)$)  rectangle ($(sincnet.north east)+(0.3,-0.6)$);
        
        \node[color=RoyalBlue, font=\scriptsize] (bluelabel) at ($(lstm1.north east)+(0.1,0.2)$) {\textcolor{RoyalBlue}{$\theta_v$ : voice activity detection}};
        \node[color=BrickRed, font=\scriptsize] (redlabel) at ($(lstm3.north east)+(0.1,-0.8)$) {\textcolor{BrickRed}{$\theta_d$ : domain classification}};
        \node[color=ForestGreen, font=\scriptsize] (greenlabel) at ($(sincnet.north west)+(0.4,0.2)$) {\textcolor{ForestGreen}{$\theta_f$ : feature extraction}};
    \end{tikzpicture}
}
\caption{Proposed architecture. The network takes the raw waveform of a 2s audio chunk as input and passes it to the part responsible for extracting the features, which is based on SincNet convolutional layers \protect\cite{Ravanelli2018}. The voice activity detection branch is made of a stack of two bi-directional LSTMs, followed by three feed-forward layers. The domain classification branch is made of one uni-directional LSTM, followed by max-pooling along the time axis and one feed-forward layer outputting the probability distribution over the domains. Depending on the task, one would only use the upper branch (for regular voice activity detection), the lower branch (for domain classification), or combine both with gradient reversal (for domain-adversarial voice activity detection).}
\label{fig:architecture}
\end{figure*}

\section{End-to-end voice activity detection}
\label{sec:vad}

Voice activity detection is the task of detecting speech regions in a given audio stream or recording. It can be addressed as a sequence labeling task where the input is the sequence of handcrafted or learnt feature vectors $\boldsymbol{x} = \{x_1, x_2, \ldots, x_T\}$ and the expected output is the corresponding sequence of labels $\boldsymbol{y} = \{y_1, y_2,\ldots, y_T\}$ where $y_{t} = 0$ if there is no speech at time step $t$ and $y_t = 1$ if there is. 
Because processing long audio files of variable lengths is neither practical nor efficient, we rely on shorter fixed-length sub-sequences for both training and inference. 

At training time, fixed-length sub-sequences~\{$\boldsymbol{x_i}\}_{i=1}^N$ are drawn randomly from the training set to form mini-batches of size $N$. 

As depicted in Figure~\ref{fig:architecture}, we propose to work directly from the waveform in an end-to-end manner, and jointly train the feature extraction and sequence labeling steps.
The upper branch is trained to minimize the cross-entropy loss, using standard gradient back-propagation to update the weights of the feature extraction network $\theta_{f}$ and the voice activity detection network $\theta_{v}$ (using abusive notation to highlight function composition):

\begin{equation}
\mathcal{L}_{v}(\theta_{f}, \theta_{v}) = - \dfrac{1}{N} 
\sum_{i=1}^{N} \sum_{t=1}^{T} \boldsymbol{{ y_{i}}_{t}} \cdot \log
\theta_{v} \left(\theta_{f} \left(\boldsymbol{x_{i}}\right)\right)_{t}
\label{eq:speech_loss}
\end{equation}

At test time, audio files are processed using overlapping sliding sub-sequences of the same length as the one used in training. 
For each time step $t$, this results in several overlapping sequences of prediction scores, which are averaged to obtain the final score. Finally, time steps with prediction scores greater than a tunable threshold $\sigma$ are marked as speech.

\section{Domain-adversarial training}
\label{sec:advtraining}
In this section, we explore how to build a feature extraction network $\theta_f$ less sensitive to domain variability. Our solution consists of adding an additional branch $\theta_d$ trained in an adversarial fashion, such as proposed in \cite{ganin2016dann}.

The corresponding lower branch in Figure~\ref{fig:architecture} is trained in conjunction with the rest of the network to minimize the mean squared error loss, MSE (using the same abusive notation as above): 
\begin{equation}
\mathcal{L}_{\text{d}}( \theta_{f},  \theta_{d}) = \dfrac{1}{N} \sum_{i=1}^{N} \norm{d_i - \theta_{d} \left(\theta_{f} \left(\boldsymbol{x_i}\right)\right)}_2^2
\label{eq:domain_loss}
\end{equation}
\noindent where $d_i$ is a one-hot encoding of the domain label of $\boldsymbol{x_i}$.
Instead of simply summing the two losses $\mathcal{L}_{\text{v}}$ and $\mathcal{L}_{\text{d}}$ as in a standard multi-task learning scenario, a gradient reversal layer~\cite{ganin2015unsupervised} is added in front of the domain classification branch. The back-propagation update rules become:
\begin{gather*}
 \theta_{v} \leftarrow  \theta_{v} - \epsilon \frac{\partial \mathcal{L}_v}{\partial  \theta_{v}} ~~~~~~~~~~~~
 \theta_{d} \leftarrow  \theta_{d} - \epsilon \frac{\partial \mathcal{L}_d}{\partial  \theta_{d}} \\
\theta_{f} \leftarrow  \theta_{f} - \epsilon \left( \frac{\partial \mathcal{L}_v}{\partial \theta_{v}} {\color{red} \boldsymbol{- \lambda}} \frac{\partial \mathcal{L}_d}{\partial  \theta_{d}} \right)
\end{gather*}

\noindent where $\epsilon$ is the learning rate and $\lambda$ is a scaling factor controlling the importance of the domain classification loss $\mathcal{L}_{\text{d}}$ with respect to the main voice activity detection loss $\mathcal{L}_{v}$.

Contrary to previous studies that assume (unlabeled) target-domain data availability during training on a (labeled) source domain~\cite{ganin2016dann}, we do not impose this restriction: we do not rely on any labeled or unlabeled target domain data during training. However, we do assume that the number and variability of domains covered by the (labeled) training set is sufficiently large. The intuition behind this statement is that the adversarial branch will reduce the space of possible feature extraction functions $\theta_{f}$ towards a solution that is less sensitive to domain variability; hence, making the whole voice activity detection branch $\theta_{v} \circ \theta_{f}$ robust to a new unseen domain (and removing the need to re-train the model for every new target domain).

\section{Experiments}
\label{sec:experiments}

\subsection{Dataset}

We perform all experiments on the single-channel subset of the DIHARD dataset~\cite{ryant2019second}. 
It contains approximately 47 hours of audio recordings, originally divided into a development set of about 24 hours of audio, and a test set of 23 hours. No training set is provided. For the purpose of this paper, we split the official development set in two parts: two thirds (16 hours, 126 files) serve  as training set, the other third (8 hours, 66 files) becomes our development set. The test set (23 hours, 194 files) remains unchanged. For each subset, files are split evenly into 11 different domains covering a wide range of conditions:
audio books, broadcast interview, child language, clinical, courtroom, map task, meeting, restaurant, socio-linguistic field recordings, socio-linguistic lab recordings, and web video. 

\subsection{Evaluation metric} 

We use the detection error rate, such as implemented in {\small \texttt{pyannote.metrics}}~\cite{pyannote.metrics}, to evaluate our systems: 
$$\text{detection error rate} = \dfrac{\text{false alarm} + \text{missed detection}}{\text{total}} $$ 
\noindent where $\text{false alarm}$ is the duration of non-speech incorrectly classified as speech, $\text{missed detection}$ is the duration of speech incorrectly classified as non-speech, and $\text{total}$ is the total duration of speech in the reference.

\subsection{Implementation details}

Figure~\ref{fig:architecture} depicts the architecture used in all experiments. For SincNet, we use the configuration proposed by the authors of the original paper~\cite{Ravanelli2018}.All long short-term memory (LSTM) and inner feed-forward (FF) layers have a size of 128 and use \emph{tanh} activations. 
The learning rate is controlled by a cyclical scheduler \cite{cyclical}, each cycle lasting for 21 epochs. All models have been trained with a batch of size 64. Data augmentation is applied directly on the waveform using  additive  noise extracted  from  the  MUSAN  database \cite{musan} with a random target signal-to-noise ratio ranging from 10 to 20 dB.

\subsection{Evaluation protocol}

For each experiment, the neural network is trained for up to 300 epochs on the training set. The development set is used to choose the actual epoch and detection threshold $\sigma$ that minimize the detection error rate.
We apply those optimal hyper-parameters and report corresponding performance on the test set. Note that, depending on the experiment, the test set domains might have or not been seen during training and development: the latter is marked as \emph{out-domain} in Table~\ref{tab:davad_perf}.

\section{Results}

\subsection{End-to-end voice activity detection}

\begin{table}
    \centering
    \caption{Evaluation of voice activity detection models, in terms of detection error rate (DetER \%), false alarm (FA \%), and missed detection (Miss \%) rates. Results on the development set are reported using small font size.  We report two variants: the first one is based on handcrafted features (MFCCs) and the other one is an end-to-end model processing the waveform directly.} 
    \begin{tabular}{|l|rr|rrrr|}
        \hline
        & \multicolumn{2}{c|}{DetER \%} & \multicolumn{2}{c|}{FA \%} & \multicolumn{2}{c|}{Miss \%} \\
        \hline
        Baseline \cite{Diez2019, ryant2019second} & 11.2 & & 6.5 & & 4.7 & \\
        \hline
        MFCC~\cite{Gelly2018} & 10.5 & {\scriptsize{10.0}} & 6.8 & {\scriptsize{5.4}} & 3.7 & {\scriptsize{4.6}} \\
        Waveform & \textbf{9.9} & {\scriptsize{9.3}} & 5.7 & {\scriptsize{3.7}} & 4.2 & {\scriptsize{5.6}} \\
        \hline
    \end{tabular}
    \label{tab:vad}
\end{table}

Table~\ref{tab:vad} summarizes the performance of the proposed end-to-end voice activity detection approach and compares it to a variant where handcrafted features (MFCCs) are used in place of trainable ones, and to the winning submission~\cite{Diez2019} of the Second DIHARD challenge~\cite{ryant2019second}. The proposed approach outperforms both of them by a significant margin, reaching a detection error rate of 9.9\%.

\subsection{Domain classification}

Before diving into domain-adversarial training, we checked whether the domain classification branch of our model was indeed able to discriminate between domains. Hence, we trained parts of the network that correspond to the domain classification branch, that are parameters $\theta_{f}$ and $\theta_{d}$ such as depicted in Figure \ref{fig:architecture}. These parameters were trained to classify domains of 2 seconds-long sub-sequences. 

\begin{figure}[h]
\includegraphics[width=\linewidth]{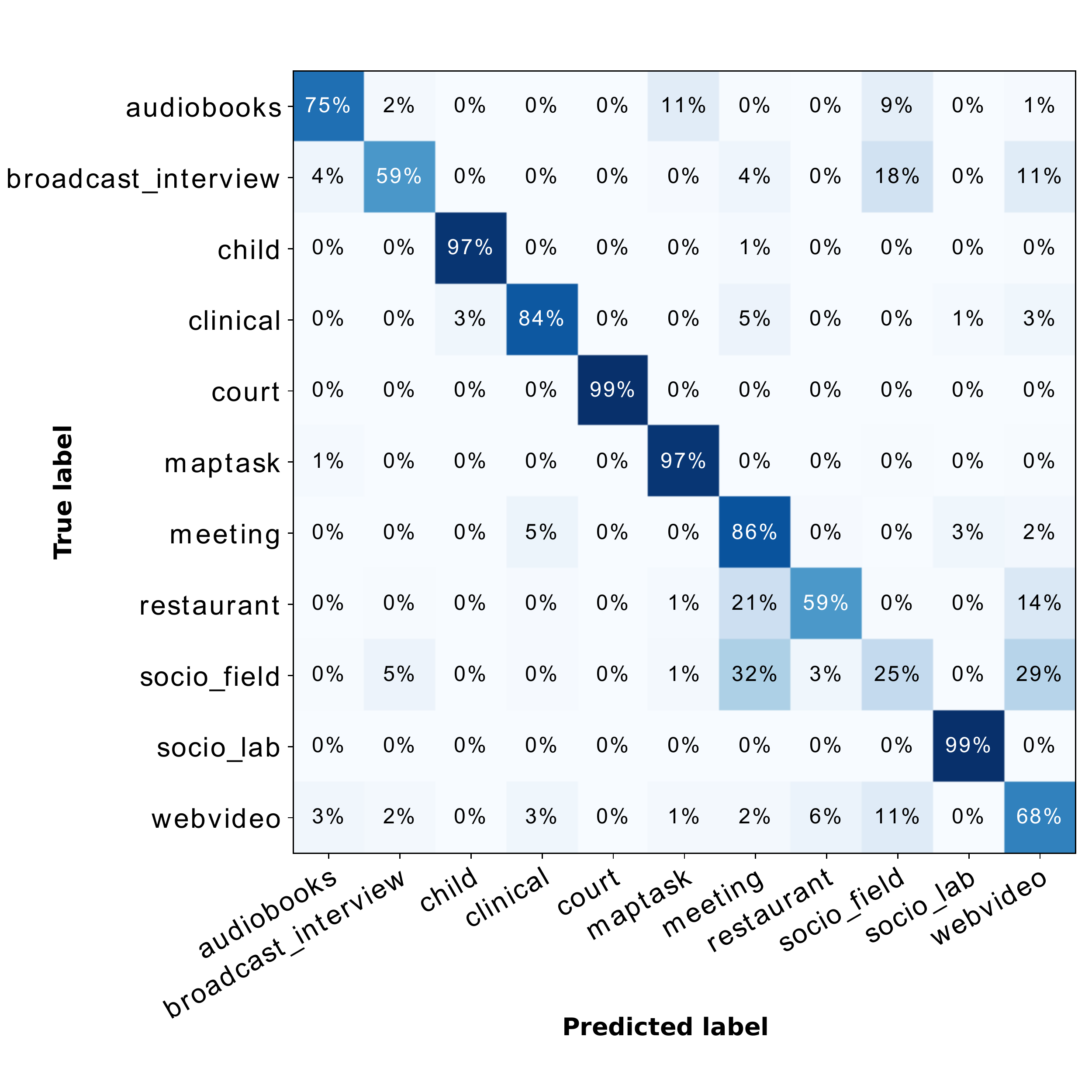}
\caption{Confusion matrix of domain predictions (columns) and true domains (rows) on 2 seconds-long sub-sequences of audio. Results are obtained on the test set.}
\label{fig:confmat}
\end{figure}

Figure \ref{fig:confmat} shows results obtained by this domain classification branch. Overall, the classifier obtained an accuracy of 77\% on the test set. The model showed reasonably good performances for most domains. Worst performance was obtained for the \textit{socio field} sub-sequences that were classified as belonging to the \textit{meeting} domain in 32\% of the cases, the \textit{webvideo} domain  in 29\% of the cases, the \textit{broadcast interview} domain in 5\% of the cases, and the \textit{socio field} domain in only 25\% of the cases.

We tried different positions for the domain classification branch (after SincNet, and after the first or the second LSTM layer) but no differences were observed, neither in terms of classification performance, nor in terms of detection error rate on the speech activity detection task.

\subsection{Domain-adversarial voice activity detection}

\begin{table*}
    \centering
    \caption{Comparison of regular and domain-adversarial voice activity detection. Results are calculated on the test set for $\lambda = 1$. Line \textbf{A} corresponds to 11 models trained, tuned and tested separately on each of the 11 domains of the dataset. Lines \textbf{C} and \textbf{B} correspond to models trained, tuned and tested on all domains at once (with or without the adversarial branch, respectively). Line \textbf{E} and \textbf{D} correspond to models trained and tuned on 10 out of the 11 domains, and tested on the left-out domain (with or without the adversarial branch, respectively). When there is one model per domain (\textbf{A}, \textbf{D} and \textbf{E}), performances of each model for their respective domain have been aggregated across all of the domains.}
    \begin{tabular}{|l|l|c|c|c|c|c|c|}
        \hline
         & \multicolumn{2}{|c|}{Training \& development} & \multicolumn{2}{c|}{Inference} & \multicolumn{3}{c|}{Metric}\\
        \hline
        & \# domains & adversarial & \# domains & out-domain & DetER & FA & Miss \\
        \hline
        \textbf{A} & $1$ &  & $1$ & & 11.3 & 6.8 & 4.5\\
        \hline
        \textbf{B} & $N$ &  & $N$ & & 9.9 & 5.7 & 4.2 \\
        \textbf{C} & $N$ & \checkmark & $N$ & & 10.1 & 6.1 & 4.0 \\
        \hline
        \textbf{D} & $N-1$ &  & $1$ & \checkmark & 13.4 & 9.0 & 4.4\\
        \textbf{E} & $N-1$ & \checkmark & $1$ & \checkmark & 11.8 & 7.6 & 4.2\\
        \hline
        
    \end{tabular}
    \label{tab:davad_perf}
\end{table*}

Returning to the speech activity detection task, Table ~\ref{tab:davad_perf} shows performances of our domain-adversarial architecture on target domains that have been seen, or not, during the training phase. 

Unsurprisingly, lines \textbf{A} and \textbf{D} show that the performances are always better when models are tested on target domains that have been seen during training. It is well-known that supervised methods are sensitive to the domain mismatch problem. In our case, this domain mismatch leads to an increase of the detection error rate of 2.1\%.

Lines \textbf{A} and \textbf{B} indicate that adding more annotated data in the training set, even though they come from a different domain, leads to a significant performance boost. This translates into a decrease of 1.4\% of the detection error rate.

The models trained adversarially on the domain classification task show a 12\% relative improvement compared to the model trained in the single-task learning setup (lines \textbf{D} and \textbf{E}).

Moreover, a comparison between lines \textbf{A} and \textbf{E} shows us that adversarially trained model performances are almost as good as models trained on a single domain. This constitutes a demonstration of the viability of the adversarial strategy for using models on unseen, unlabeled or slightly annotated data. 

Finally, lines \textbf{B} and \textbf{C} indicate that adversarially training on the classification task does not help to improve performances on domains that have been already seen during the training. This result led us to think that the domain can contain useful information to identify speech and non-speech segments, probably for further refining boundaries between speech and noise.

\begin{figure}[h]
\includegraphics[width=\columnwidth]{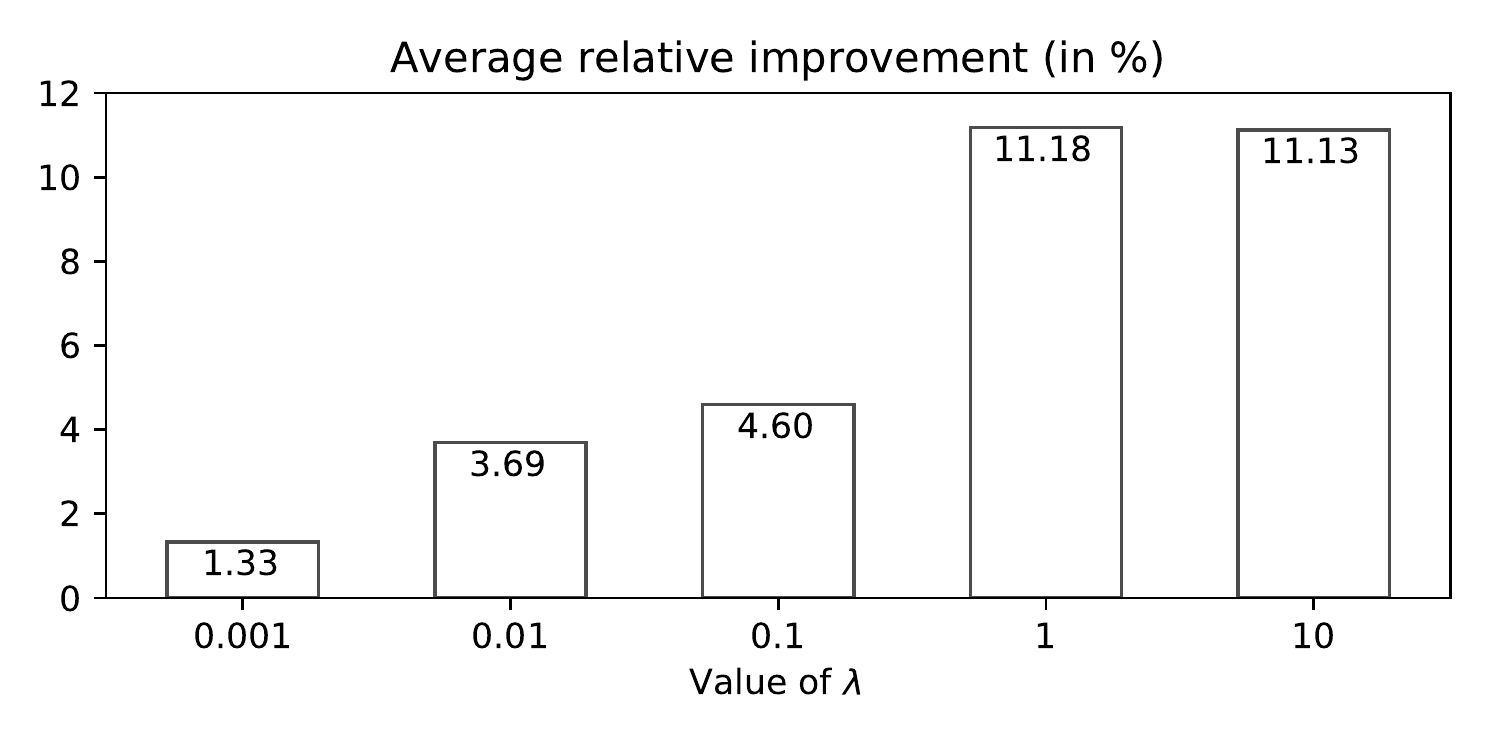}
\caption{Relative improvement (in \%) in terms of detection error rate for different values of $\lambda$ compared to the single-task voice activity detection in the leave-one-out setup.}
\label{fig:lambda_plot}
\end{figure}

In Figure \ref{fig:lambda_plot}, we show relative improvement, aggregated across domains, in terms of detection error rate for different values of $\lambda$, the weight associated to the domain classification loss.
Higher is the value of $\lambda$, higher is the average relative improvement, until it gets eventually too high and too much attention is given to the domain classification task. Best results are obtained for $\lambda = 1$ and $\lambda = 10$ that decrease, on average, the detection error rate by $11.15 \%$.

\section{Reproducible research}

All the code has been implemented (and integrated into) using  {\small \texttt{pyannote.audio}} \cite{Bredin2020}, a python toolkit to build neural networks for the speaker diarization task. A Github page\footnote{https://github.com/hbredin/DomainAdversarialVoiceActivityDetection} provides instructions to reproduce results, along with additional results such as a per-domain analysis as well as ready-to-use pretrained models.

\section{Conclusions}
\label{sec:conclusion}

This paper explores the learning of filters using the SincNet model \cite{Ravanelli2018}, in conjunction with the use of domain-adversarial neural networks \cite{ganin2016dann} for explicitly extracting domain-independent features. 

We show that end-to-end voice activity detection leads to a significant improvement compared to models based on handcrafted features. 

Furthermore, when applied on unseen domains, the domain-adversarial multi-task learning strategy greatly improves the performances compared to the standard VAD model which does not use the domain information. Therefore, on the voice activity detection task, it seems that the strategy of muting the domain information appears as viable for extracting robust features that generalize well to new unseen domains. This method potentially reduces the need for labelled data and can improve performances on downstream tasks such as the speaker diarization or the speech recognition task that require robust voice activity detection systems. 

Finally, we provide a fully reproducible open-source pipeline that can be easily adapted to other datasets as well as ready-to-use pretrained models.

\section{Acknowledgements}

The research reported here was conducted at the 2019 Frederick Jelinek Memorial Summer Workshop on Speech and Language Technologies, hosted at L'\'Ecole de Technologie Sup\'erieure (Montreal, Canada) and sponsored by Johns Hopkins University with unrestricted gifts from Amazon, Facebook, Google, and Microsoft. This work also benefited from the support of the Analyzing Child Language Experiences around the World (ACLEW) collaborative project ANR-17-CE28-0007 LangAge, ANR-16-DATA-0004 ACLEW, ANR-17-EURE-0017, ANR-16-CE92-0025 PLUMCOT of the French National Research Agency (ANR). We would like to thank Neville Ryant for providing the speaker diarization output of the winning submission to DIHARD 2019.

\vfill\pagebreak

\bibliographystyle{IEEEtran}
\bibliography{refs}

\end{document}